\documentstyle[psfig,raulbib,times]{mn}

\begin{document}

\newcommand{\be}{\begin{equation}}
\newcommand{\ee}{\end{equation}}
\newcommand{\ba}{\begin{eqnarray}}
\newcommand{\ea}{\end{eqnarray}}
\newcommand{\mat}{{\bf}}
\newcommand{\nn}{\nonumber\\}
\newcommand{\bA}{{\bf A}}
\newcommand{\bE}{{\bf E}}
\newcommand{\bB}{{\bf B}}
\newcommand{\bC}{{\bf C}}
\newcommand{\bF}{{\bf F}}
\newcommand{\bS}{{\bf S}}
\newcommand{\bfa}{{\bf a}}
\newcommand{\bb}{{\bf b}}
\newcommand{\bg}{{\bf g}}
\newcommand{\bj}{{\bf j}}
\newcommand{\bk}{{\bf k}}
\newcommand{\bn}{{\bf n}}
\newcommand{\bp}{{\bf p}}
\newcommand{\br}{{\bf r}}
\newcommand{\bu}{{\bf u}}
\newcommand{\bv}{{\bf v}}
\newcommand{\bx}{{\bf x}}
\newcommand{\by}{{\bf y}}
\newcommand{\bbmu}{\mbox{\boldmath $\mu$}}
\newcommand{\dd}{{\partial}}
\newcommand{\ddt}{{\partial\over \partial t}}
\newcommand{\lnL}{\ln{\cal L}}
\newcommand{\bbtheta}{\mbox{\boldmath $\theta$}}
\def\gs{\mathrel{\raise1.16pt\hbox{$>$}\kern-7.0pt 
\lower3.06pt\hbox{{$\scriptstyle \sim$}}}}         
\def\ls{\mathrel{\raise1.16pt\hbox{$<$}\kern-7.0pt 
\lower3.06pt\hbox{{$\scriptstyle \sim$}}}}         

\title[Recovering physical parameters from galaxy spectra using MOPED]
{Recovering physical parameters from galaxy spectra using MOPED}
\author[Reichardt, Jimenez \& Heavens]{Christian Reichardt$^1$, Raul
Jimenez$^2$ \& Alan F. Heavens$^3$ \\ $^1$ Mail Code 130--33,
California Institute of Technology, Pasadena, CA 91125 USA
(cr@its.caltech.edu)\\ $^2$ Department of Physics and Astronomy,
Rutgers University, 136 Frelinghuysen Road, Piscataway, NJ 08854--8019
USA (raulj@physics.rutgers.edu)\\ $^3$ Institute for Astronomy,
University of Edinburgh, Blackford Hill, Edinburgh EH9 3HJ,
UK. (afh@roe.ac.uk)}
\maketitle

\begin{abstract}
  We derive physical parameters of galaxies from their observed spectrum,
  using MOPED, the optimized data compression algorithm of \scite{HJL00}.
  Here we concentrate on parametrising galaxy properties, and apply the method
  to the NGC galaxies in Kennicutt's spectral atlas.  We focus on deriving the
  star formation history, metallicity and dust content of galaxies.  The
  method is very fast, taking a few seconds of CPU time to estimate $\sim 17$
  parameters, and so specially suited to study of large data sets, such as the
  Anglo-Australian 2 degree field galaxy survey and the Sloan Digital Sky
  Survey.  Without the power of MOPED, the recovery of star formation
  histories in these surveys would be impractical.  In the Kennicutt atlas, we
  find that for the spheroidals a small recent burst of star formation is
  required to provide the best fit to the spectrum. There is clearly a need
  for theoretical stellar atmospheric models with spectral resolution better
  than 1\AA\, if we are to extract all the rich information that large
  redshift surveys contain in their galaxy spectra.
\end{abstract}

\section{Introduction}

Most of the information about the physical properties of galaxies comes from
their electromagnetic spectrum. It is therefore of paramount importance to be
able to extract as much physical information as possible from it. In
principle, it is straightforward to determine physical parameters from an
individual galaxy spectrum. The method consists of building synthetic stellar
population models which cover a large enough range in the parameter space and
then use a merit function (typically $\chi^2$ if the pixels are not
correlated) to evaluate which suite of parameters fits the observed spectrum
best. There are two obvious limitations of the above method: first, the number
of parameters that govern the spectrum of a galaxy may be very large and thus
difficult to explore fully; secondly, in the case of ongoing large redshifts
surveys which will provide us with about a million galaxy spectra, it will be
computationally very expensive (and possibly intractable for redshift surveys
like the 2dF and SDSS) to apply a brute-force $\chi^2$ analysis to each
individual spectrum which itself may contain of the order of $10^3$ data
points.

A less obvious route to tackle the high computational requirement is to
compress the original data set, giving more weight to those pixels in the
spectrum that carry most information about a given parameter.  In this paper
we show how this can be done in an optimal way.  It is worth remembering that
data compression is commonly applied to galaxy spectra, either by the
instrument, through the use of photometric filters, or in the interpretation,
by concentrating on specific spectral features and ignoring others.  Not
surprisingly, this empirical data compression is not optimal since it is ad
hoc.  For example, the photometric $B$ filter alone is not optimal to recover
the age of a galaxy. On the other hand, more sophisticated and non-empirical
methods have been proposed for extracting information from galaxy spectra,
some of them as old as the Johnson's filter system.  Many of these are based
on Principal Component Analysis (PCA) or wavelet decomposition
\cite{Murtagh87,Francis92,Connolly95,Folkes96,Galaz98,Bromley98,Glazebrook98,Singh98,Connolly99,Ronen99,Folkes99}.
PCA projects galaxy spectra onto a small number of orthogonal components. The
weighting of each component corresponds to its relative importance in the
spectrum. However while these components appear to correlate reasonably well
with physical properties of galaxies, their interpretation is difficult since
they do not have known, specific physical properties -- they can be amalgams
of different properties. To interpret these components, we have to return to
model spectra and compare them with the components \cite{Ronen99}. This is a
disadvantage of PCA since one important goal of the analysis is to study the
evolution of the physical properties which dramatically affect galaxy spectra,
such as the age, metallicity, star formation history or dust content.  It is
important to recognise that PCA can play an important role if there is no
underlying model for how the data should behave.  If such a model exists, then
one can do better by using projections of the data which are designed to give
the parameters of the model as accurately as possible.

An optimal parameter--extraction method, which we term
MOPED\footnote{The MOPED algorithm has a patent pending}
(Multiple Optimised Parameter Estimation and Data compression) was
developed in \scite{HJL00}. The purpose of this paper is to apply
the MOPED method to a specific set of observations, to estimate
their physical parameters and thus demonstrate the usefulness of
the approach. We also want to demonstrate with a specific example
the massive speed up factor that the method provides, making
computationally-intensive problems into much more accessible
ones. The outline of the paper is at follows: in section II we
briefly describe the method used; in section III we discuss the
problems involved in extracting physical parameters from galaxy
spectra: how to choose the best parametrisation for the star
formation history; in section IV we derive physical parameters
for the Kennicutt atlas and present the best fitting models in an
appendix.

\section{The method}

The MOPED algorithm was presented in detail in \scite{HJL00}. Here
we briefly recall the main ingredients and steps of the method.

The main idea of the method is that, in practice, some of the data may
tell us very little about the parameters we are trying to estimate,
either through being very noisy, or through having no sensitivity to
the parameters. So in principle, we may be able to throw away some
data without losing much information about the parameters. It is
obvious that simply throwing away some of the data is not in general
optimal; it will usually lose information. On the other hand, by
constructing linear combinations of the data we might do better and
then we can throw away the linear combinations which tell us least.
In fact one can do much better than this.  Providing the noise has
certain properties, one can reduce the size of the dataset down to a
handful of linear combinations -- one for each parameter -- which
contain as much information as the entire dataset.  It is by no means
obvious that this can be done.

Given a set of data {\bf x} (in our case the spectrum of a galaxy) which
includes a signal part ${\bbmu}$ and noise ${\bf n}$, i.e. $\bx = \bbmu +
\bn$, the idea then is to find weighting vectors ${\bf b}_m$ such that $y_m
\equiv {\bf b}_m^{t} {\bf x}$ contain as much information as possible about
the parameters (star formation rates, metallicity etc.).  These {\it numbers}
$y_m$ are then used as the data set in a likelihood analysis.

In \scite{HJL00} an optimal and lossless method was found to calculate ${\bf
  b}_m$ for multiple parameters (as is the case with galaxy spectra).  The
definition of lossless here is that the Fisher matrix at the
maximum likelihood point (see \pcite{TTH97}) is the same whether
we use the full dataset or the compressed version.  The Fisher
matrix gives a good estimate of the errors on the parameters,
provided the likelihood surface is well described by a
multivariate Gaussian near the peak.  We find that the method is
lossless provided that the noise is independent of the
parameters.  This is not exactly true for galaxy spectra, owing
to the presence of a shot noise component from the source
photons.  However, the increase in parameter errors is very small
in this case (see \pcite{HJL00}).  The weights required are
\begin{equation}
\bb_1 = {\bC^{-1} \bbmu_{,1}\over \sqrt{\bbmu_{,1}^t
\bC^{-1}\bbmu_{,1}}}
\label{Evector1}
\end{equation}
and
\begin{equation}
\bb_m = {\bC^{-1}\bbmu_{,m} - \sum_{q=1}^{m-1}(\bbmu_{,m}^t
\bb_q)\bb_q \over
\sqrt{\bbmu_{,m} \bC^{-1} \bbmu_{,m} - \sum_{q=1}^{m-1}
(\bbmu_{,m}^t \bb_q)^2}}.
\label{bbm}
\end{equation}
where a comma denotes the partial derivative with respect to the
parameter $m$ and $C$ is the covariance matrix with components
$C_{ij}=\langle n_in_j\rangle$. $m$ runs from 1 to the number of
parameters $M$, and $i$ and $j$ from 1 to the size of the dataset
(the number of flux measurements in a spectrum).  To compute the
weight vectors requires an initial guess of the parameters.  We
term this the fiducial model.

The dataset $\{y_m\}$ is orthonormal: i.e. the $y_m$ are
uncorrelated, and of unit variance. The new likelihood is easy to
compute (the $y_m$ have mean $\langle y_m \rangle = \bb_m^t
\bbmu$), namely: \be \ln{\cal L}(\theta_\alpha) = {\rm constant}
- \sum_{m=1}^{M} {(y_m-\langle y_m\rangle)^2\over 2}. \ee Further
details are given in \scite{HJL00}.

It is important to note that if the covariance matrix is known for a large
dataset (e.g. a large galaxy redshift survey) or it does not change
significantly from spectrum to spectrum, then the $\langle y_m \rangle$ need be
computed only {\em once} for the whole dataset, thus with massive speed up
factors in computing the likelihood as will be shown in sections 3 and 4.
Note that the $y_m$ are only orthonormal if the fiducial model coincides with
the correct one.  In practice one finds that the recovered parameters are
almost completely independent of the choice of fiducial model, but one can
iterate if desired to improve the solution.

\subsection{Estimating errors in the recovered parameters}

If we have $M$ parameters, we can get the marginal error on each of
the parameters from the Fisher matrix
\begin{equation}
F_{ij} \equiv -\left\langle {\partial^2\ln{\cal L}\over \partial
\theta_i \partial\theta_j}\right\rangle
\end{equation}
where $i$ and $j$ run from 1 to $M$ (Note the size of the dataset appears
nowhere here - it is only relevant in computing $\lnL$).  The averaging is
over many realisations.

The {\em conditional} error on $\theta_i$ (fixing all other
parameters) is just $(F_{ii})^{-1/2}$ (no summation).  This is
not too relevant as we wish to estimate all the parameters
simultaneously. More relevant then is the {\em marginal} error,
which is
\begin{equation}
\sigma_i = \sqrt{(F^{-1})_{ii}}.
\end{equation}
So the procedure to follow is to estimate $F_{ij}$ from the
likelihood surface near the peak, invert the $M\times M$ Fisher
matrix and use the diagonal components of the inverse matrix to
assign marginal errors.  All this assumes that the likelihood
surface is well-approximated by a multivariate Gaussian, not just
at the peak, but also far enough down the likelihood hill (so
$\ln{\cal L}$ drops by around unity).

\subsubsection*{Fisher matrix estimation}

Assuming that the maximisation finds the maximum precisely, at position
$\bbtheta_0$ (an $M-$dimensional vector in parameter space), with
value $\lnL_0$, then a Taylor expansion of $\lnL$ around the maximum
gives
\begin{equation}
\lnL(\bbtheta_0+\Delta\bbtheta) \simeq \lnL_0 + {1\over
2}{\partial^2\ln{\cal L}\over \partial \theta_i
\theta_j}\Delta\theta_i \Delta\theta_j
\end{equation}
where the summation convention is assumed.  It will not apply for
the rest of this section.

We estimate the diagonal components of the second derivatives by keeping all components of
$\bbtheta$ constant apart from a single $\theta_i$.
\begin{eqnarray}
F_{ii}&  \simeq& \\\nonumber
& & -{1\over \Delta\theta_i^2} \left[\lnL(\bbtheta_0 +
\Delta\theta_i {\bf e}_i) +  \lnL(\bbtheta_0 -
\Delta\theta_i {\bf e}_i) - 2 \lnL_0\right].
\label{Fii}
\end{eqnarray}

Similarly, the off-diagonal terms are estimated from
\begin{eqnarray}
F_{ij} & \simeq & {-1\over 2 \Delta\theta_i \Delta\theta_j}  \left[
\right.
\\\nonumber
& &\lnL(\bbtheta_0 + \Delta\theta_i {\bf e}_i + \Delta\theta_j {\bf e}_j)+
\\\nonumber
& &\lnL(\bbtheta_0 - \Delta\theta_i {\bf e}_i -
\Delta\theta_j {\bf e}_j)- \\\nonumber
& &\lnL(\bbtheta_0 - \Delta\theta_i {\bf e}_i +
\Delta\theta_j {\bf e}_j)- \\\nonumber
& &\lnL(\bbtheta_0 + \Delta\theta_i {\bf e}_i -
\Delta\theta_j {\bf e}_j)
\left .\right].
\label{Fij}
\end{eqnarray}
The above procedure is not computationally expensive - for $M=17$
we need $17\times 16/2$ off-diagonal terms, and $17$ diagonal
terms.  This requires about 600 likelihood evaluations.  The
marginal errors are estimated from $\sigma_i \simeq
1/\sqrt{(F^{-1})_{ii}}$.

\section{Determining physical parameters from galaxy spectra}

\subsection{The problem}

Our aim is to determine the star formation history, metallicity
and dust evolution of the stellar population of a galaxy from its
spectrum. We wish to do this for galaxy spectra that typically
contain thousands of data points. Additionally, current large
surveys contain on the order of $10^6$ galaxies. It is
computationally very expensive to compute a ``brute-force''
likelihood analysis with all the data points. For example, for a
spectrum with $10^3$ points for which we want to determine 17
parameters, it takes about 1.5 hours of CPU time on an Athlon
1Ghz machine to find the minimum of the likelihood surface.  With
$10^6$ spectra, it would take about 170 years, and if the
theoretical models change, it would take as long again.  Clearly,
the brute force approach is too slow by a factor of several
hundred. As we show below, we can estimate 17 parameters of a
galaxy with MOPED in about 20 seconds.
\begin{figure}
\centerline{\psfig{figure=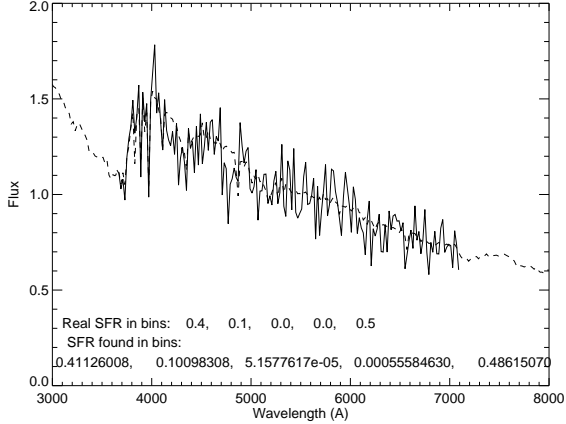,height=6cm,angle=0}}
\caption{(Solid line) Simulated spectrum with noise ($S/N=10$).
  This spectrum was created with 5 time bins, each of height indicated in the
  plot. The metallicity in each bin was kept constant at the solar value. The
  dashed line depicts the recovered spectrum using MOPED. The recovered values
  for the SF in each bin are also given, showing good agreement.}
\label{rec1}
\end{figure}
\begin{figure}
\centerline{\psfig{figure=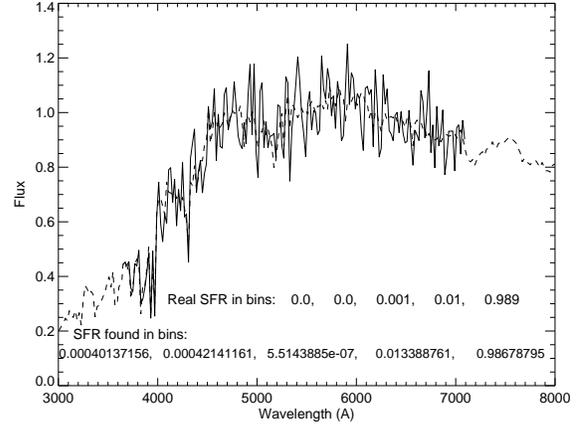,height=6cm,angle=0}}
\caption{Same as Figure~\ref{rec1} but for a different SFH, in this case the
  agreement is also reasonable.}
\label{rec2}
\end{figure}

\begin{figure}
\centerline{\psfig{figure=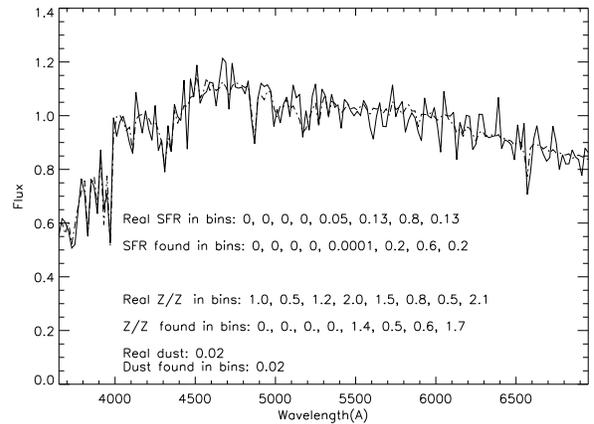,height=6cm,angle=0}}
\caption{Same as Figure~\ref{rec1} but for a different SFH and arbitrary
  metallicity in the bins.  With varying metallicity, higher signal-to-noise is required to
  obtain unambiguous correct recovery with this short, low-resolution spectrum.}
\label{rec3}
\end{figure}

\subsection{Choosing the optimal parametrisation}

Which parametrisation should we choose in order to determine the
star formation history (SFH) of a galaxy? The common procedure in
the literature is to {\it assume} that the SFH proceeded as a
decaying exponential law with one parameter, while more
sophisticated modifications allow for the presence of a burst (or
two) at a given time. Ideally one would prefer to avoid any
assumptions about the form of the SFH.  In fact, star formation
in galaxies takes place in giant molecular clouds which are short
lived (about $10^7$ years) due to the explosion of SN. It is
therefore clear that we could parametrise the SFH in a rather
model--independent way by dividing time into bins of width
smaller than $10^7$ years and using the height of each bin as the
parameter.  Of course, for galaxies as old as $10^{10}$ years this
approach becomes useless, since one can hardly deal with over a
thousand parameters.  We therefore choose a different approach.
We divide time into a fixed number of wider bins, and choose their
positions and widths according to the following scheme.  We
consider bursts of star formation at the beginning and end of
each time bin (at a fixed metallicity), and require that the
maximum fractional difference in the final spectrum is the same
for each bin.  The maximum difference depends on how many bins we
wish to consider.  To illustrate: for a galaxy whose age is 14
Gyr with a maximum error in the height of each bin of a factor of
2 (100\% error), the optimal bin boundaries will be at 0.03,
0.23, 0.42, 0.66, 0.92, 2.56, 6.33 and 14 Gyr (considering fixed
solar metallicity).  Further exploration with different
metallicities showed us that the bin boundaries differ by a small
amount. In fact the bin boundaries are very close to equally
spaced bins in logarithmic space -- which for the above case are:
0.02, 0.05, 0.13, 0.33, 0.84, 2.15, 5.49 and 14.0 Gyr.  For
convenience, we choose equally-spaced logarithmic bin boundaries
in what follows. Therefore, we are faced now with 8 parameters to
determine the SFH. Note that we have not made any specific
assumption about the actual shape of the SFH.  Within this
framework metallicity is extremely easy to parametrise since to
each bin we simply assign an extra parameter which is the
metallicity of the {\it stars} formed in that bin.

Another parameter to consider is the dust content of the galaxy.
The importance of dust is being recognised (e.g.
\pcite{JPDBJM00}), but modelling is difficult and currently much
less sophisticated than the stellar population modelling. Since
the purpose of this paper is only to illustrate the usefulness of
the method, we will not develop very sophisticated models for the
process of dust emission and absorption. We use the \scite{C97}
parametrisation as sufficient to describe the major effect of
dust absorption on the integrated light of galaxies. The Calzetti
model depends only on one parameter: the amount of dust in the
galaxy. We will therefore use this single parameter to describe
the global effect of dust. Note that as more realistic dust
parametrisations are proposed (e.g. \scite{JPDBJM00,CF00}), the
search parameter space increases and data compression methods
become increasingly necessary.

Emission lines are not naturally predicted in the spectrophotometric
models.  They could be included at the expense of a larger parameter
space to explore.  An alternative approach, which we follow here, is
not to include the emission lines in the analysis at all.  Thus we
will not expect to fit the emission lines, but only the continuum and
absorption lines.  Similarly, we could include one (or more)
parameters characterising the velocity dispersion of the galaxy.  With
the coarse spectral resolution of the models, we have chosen not to do
this here.

\subsection{Recovery of parameters of simulated spectra}

In what follows, and for convenience, we will use the set of synthetic stellar
population models developed in \scite{JPMH98}. We emphasize that the choice of
models is not crucial at any point in our argument. Furthermore, any suit of
models can be used in principle.  It is worth noting that given the low
spectral resolution of the models (20 \AA) we are rather limited in our
capacity to extract information from the spectrum. Also, and since we are
aiming (in this particular paper) to extract parameters from the \scite{k92}
atlas, we will limit the wavelength coverage between 3800 and 7000 \AA. Note
that this is very restrictive since with a larger spectral coverage and, more
importantly, better spectral resolution of the models, we would be able to
extract a larger number of parameters with a smaller error (see section 5).

We are now in a position to test our method. We can do this by
building synthetic models with known star formation histories,
metallicities and dust and then try the parameter recovery. Fig.
\ref{rec1} shows the model spectrum (solid black line) which has
been constructed out of five bins with values: 0.4, 0.1, 0.0, 0.0
and 0.5 with fixed solar metallicity (no dust) in each bin and to
which artificial Poisson noise has been added ($S/N=10$). The
dashed line shows the best recovered model with bin values
corresponding to: 0.41, 0.10, 0.0, 0.0 and 0.49. The agreement is
remarkable.  Fig.~\ref{rec2} shows another example for constant
metallicity, again the efficiency of the method is excellent.

Encouraged by the above results we now consider a more general case where we
allow the metallicity of each bin to take an arbitrary value. We also include
dust as a free parameter. The first synthetic test case has got artificial
Poisson noise added with $S/N=10$. In this case the recovery of all 17
parameters is not optimal. The reason for this is that with such a low $S/N$
most of the information that is contained in the absorption features has
disappeared; we are left only with a continuum. Given the fact of our limited
wavelength range in this example (3800 to 7000 \AA), it is almost impossible
to break the ``famous'' age--metallicity degeneracy with only the continuum
shape. This is in stark contrast with the first test case where the
metallicity and dust where kept fixed, then it was possible with only the
continuum to derive the heights in the bins. Therefore, we now increase the
$S/N$ until we are able to recover the 17 parameters. This happens for
$S/N=20$ and is illustrated in Fig.~\ref{rec3}. It can be seen that the
parameter recovery is very good.

Given the success of the above tests we decided to apply the method to a real
sample, but before doing it we turn our attention to the physical
interpretation of the eigenvectors ${\bf b}$.

\subsection{Eigenvectors or where is the information?}

\begin{figure}
\centerline{\psfig{figure=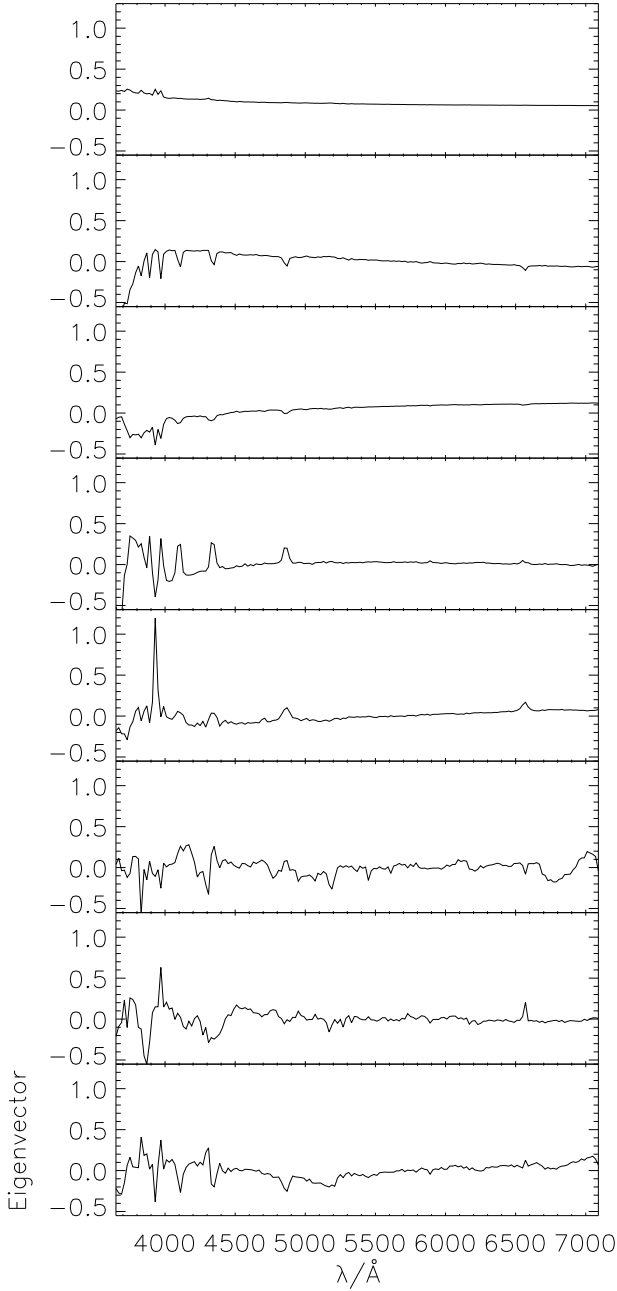,height=18cm,angle=0}}
\caption{From the top: eigenvectors {\bf b$_1$} to {\bf b$_8$}
for the SFH. The fiducial model is the same for all the
galaxies.  It corresponds to equal star formation in each bin,
solar metallicity and dust parameter=0.05.} \label{eigen}
\end{figure}

\begin{figure}
\centerline{\psfig{figure=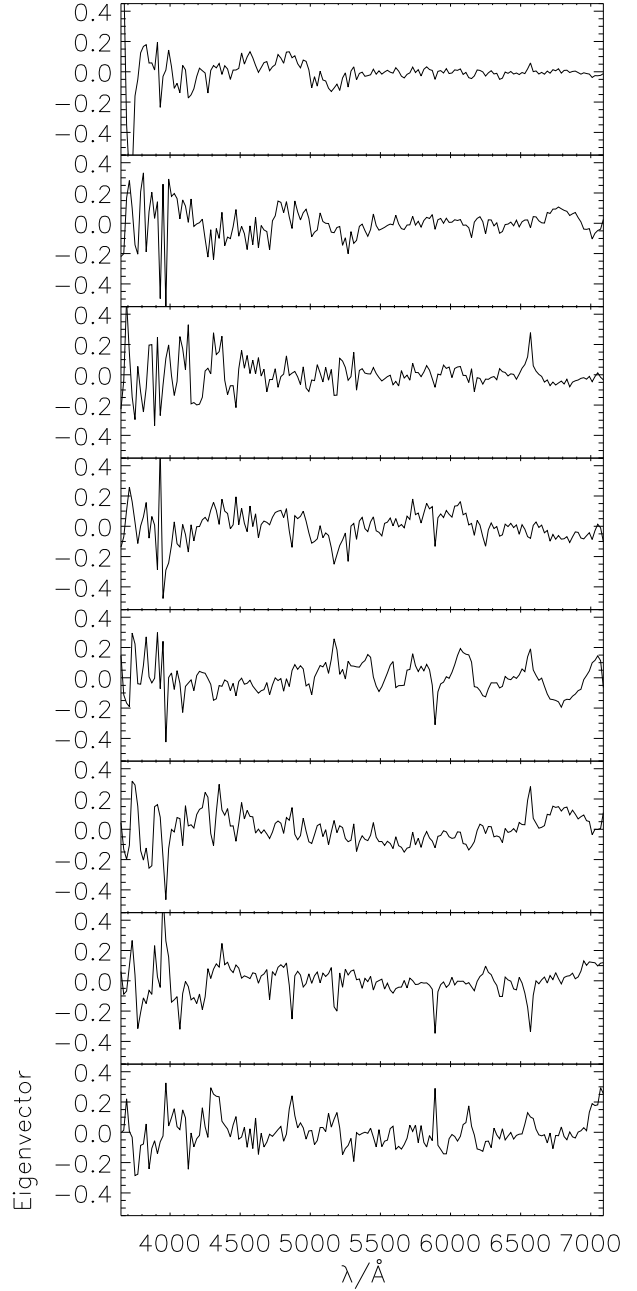,height=18cm,angle=0}}
\caption{Eigenvectors {\bf b$_9$} to {\bf b$_{16}$} for the
metallicity.} \label{eigenm}
\end{figure}

We now concentrate our attention on the physical meaning of the eigenvectors
computed from eq. (1) and (2). As shown in \scite{HJL00} the choice of the
fiducial model is not important.  One can always iterate, but this
appears to be quite unnecessary.
Figures~\ref{eigen} and \ref{eigenm} show the SFH and metallicity eigenvectors
respectively as a function of wavelength in the range 3600 to 7000 \AA\ for a
model in which we have parametrised the SFH by using 8 bins.
Figure~\ref{eigen} concerns the SFH eigenvectors for bins corresponding to :
0.02, 0.05, 0.13, 0.33, 0.84, 2.15, 5.49 and 14.0 Gyr (from top to bottom),
while the 8 panels on Figure~\ref{eigenm} are the corresponding metallicity
eigenvectors. Not surprisingly, for young ages, the age eigenvector likes to
weight most those pixels bluewards of 4500 \AA, while those redwards of 4500
\AA\ get very little weight (note that the eigenvector can always be multiplied
by $-1$). As the population ages most of the information
about the SFH becomes more distributed among wavelength, thus making it very
difficult to design a narrow band filter which would capture most of the
information. A similar situation occurs for the metallicity bins (see
Figure~\ref{eigenm}, for young ages most of the information is concentrated on
the bins bluewards of 5000 \AA\, while for older ages the information becomes
more equally distributed among different absorption features. Note that for
recovering metallicity information the absorption features are more relevant
than the continuum.

\section{Physical parameters for the spectra in the Kennicutt atlas}

In this section we apply the method to the Kennicutt atlas (\pcite{k92}).
Although the Kennicutt sample is certainly small, about 50 galaxies, the S/N
of the spectra is very high. Furthermore, the largest redshift survey
available at the moment, the 2dF, suffers from serious complications in
calibrating photometrically the continuum of the spectra \cite{MLT2dF00},
thus, for the time being, we concentrate in illustrating the method on the
small but significant atlas of observed spectra provided by \scite{k92}.  We
have re-binned the data to the same spectral resolution as the models, with
the new flux calculated as the mean flux unit wavelength in the bin.  It is
worth keeping in mind that the spectrophotometric calibration of the
\scite{k92} atlas has an error of 10\% and that our models do not include
emission lines in the spectra.  It should be stressed that despite the
excellent fits obtained to the \scite{k92} atlas, one should not
over--interpret the results since the 10\% uncertainty in the
spectrophotometric calibration affects the parameter
determination. As stated above in the paper the physical parameters we are
aiming at deriving are: the amount of star formation in the time bins chosen
-- here we chose 8, the metallicity in each bin and the global dust content,
using the \scite{C97} formula for the dust extinction. This gives a total of
17 parameters for each galaxy.  This is the most effective way to parametrise
the SF since it does not depend on any previous knowledge about its shape,
i.e. we are not assuming that it is a declining exponential or similar.  Note
that if we use the full dataset instead of the compressed data, we get the
same results (although it takes much longer (see section 5)).

Results are presented in the next 12 figures, where the best fit to the
spectrum is drawn in the left panel and the corresponding SFH is plotted with
horizontal error bars denoting the width of the bins and vertical error bars
for the marginal uncertainty in the height of the bin. The derived metallicity
for each bin is also labeled in the right panel as well as the dust with
corresponding errors.

The first result is somewhat surprising. We find that all galaxies
classified as E/S0 (NGC3245, NGC3379, NGC3516, NGC4472, NGC4648,
NGC4889, NGC5866, NGC6052) show a reasonable level of recent star
formation activity although in all cases more than 50\% of the stellar
mass is older than 5 Gyr.  NGC3245 (S0) shows recent star formation
below 7\% of the total mass as does NGC6052 (S0). NGC3379, NGC3516,
NGC4472, NGC4648, NGC4889 and NGC5866 all show a significant ( $>$
30\% of the total mass) amount of star formation in the last 3 Gyr
(Note NGC3516 has a Seyfert nucleus). Note also that the metallicities
derived are very reasonable since the final mass--weighted metallicity
for the above galaxies is about 1.5 times the solar value\footnote{due
to the fact that the
\scite{k92} sample is nearby, the slit of the spectrograph can only
sample the inner regions of the galaxy, where the metallicity is
higher.}. Note that these findings are in excellent agreement with the
recent findings about the nature of spheroids by
e.g. \scite{D+96,Spinrad+97,MAE00,DFFIK00}.  Also significant is
that the method is able to recognize current star formation, despite
the fact of not having any emission lines included in the model. For
example, for galaxies with slight H$_{\alpha}$ (like NGC1357, NGC3147,
NGC3227, NGC3921 and NGC4750) the method always shows a significant
level of recent star formation in the most recent bins, which agrees
with the star formation rate that would be estimated using the
H$_{\alpha}$ line itself (NB NGC3227 is complicated by the presence of
a Seyfert component). As expected, galaxies classified as Sa to Sc
show more recent bursts of star formation, but note that star
formation activity does not correlate with the Hubble type. In other
words, star formation history in galaxies does not proceed as a single
exponential decaying law that changes according to the Hubble type in
a monotonic manner but it is much more like a sequence of burst
events. We now turn our attention to the metallicity evolution of the
stars. The general trend is that high-metallicity stars are formed at
an early epoch, the typical values being over the solar value. On the
other hand the youngest bursts tend to have slightly sub-solar
metallicities (about half the solar value). Note that this trend is in
good agreement with predictions from infall models (e.g.
\scite{Pagel97}). These models are motivated by the fact that if all the gas
was available for consumption into stars from the very beginning (closed--box
model), then a large number of low metallicity stars should be observed now.
What is observed is that the metallicities of stars in galaxies (both disks
and ellipticals) are gaussianly distributed. An obvious solution to this
problem is not to allow all the gas to be acreted into the galaxy at the
beginning but at a slower pace. As a result of this the late forming
generations of stars will have a reservoir of fresh low metal gas which will
decrease the metallicity that would be predicted by a closed--box model, which
increases monotonically with time. Inspection of the figures in the appendix
shows that this is not the case, but that late forming generations have a
lower metallicity. Note also that where the star formation is effectively
zero, the metallicity is not recovered accurately, as one might expect.  We
note also that dust values seem to be very reasonable.

\section{Discussion and conclusions}

We have presented a fast and efficient algorithm (MOPED) to
recover physical parameters from galaxy spectra. The algorithm is
based on a data representation technique which discards most of
the data set that does not contain any information with respect
to a given parameter. By doing so we are able to speed up the
parameter determination by a large factor: typically of the order
of $N/M =$ number of data points/number of parameters. We note
here that if the noise matrix is non-diagonal, the speed-up is
much larger $\sim N^3/M$ (see Gupta \& Heavens, in preparation).
We have also applied the method to the \scite{k92} atlas and
derive star formation and metallicity histories for the NGC galaxies
in the survey. We have found that most spheroids, albeit having at
least 50\% stars formed at ages older than 5 Gyr, have had recent
star formation episodes. The quality of the spectral models does
not allow us yet to determine if those are the consequence of
recent mergers.

The method presented in section 2.1 is good at estimating the {\it
local} errors around the maximum. Unfortunately, the thing that
usually happens is that the error estimation is mostly dominated
by local maxima that are in distant regions of the parameter
space, this is our case. In order to explore this we record all
the maxima that the method finds starting from different initial
random guesses (in this case 5000) \footnote{The space parameter
is so large that one cannot compute the whole likelihood surface,
instead we used a conjugate gradient method to find the maximum of
the likelihood. In order to test that we had found the absolute
maximum we repeated the process from new random initial guesses.
We found that 400 of these guesses suffice to find the global
maximum.}.We then explore if some of these solutions are also
allowed in the $\chi^2$ sense. We have done this for all the
spectra in the \scite{k92} atlas. Our findings are as follows:
for systems with only absorption features the number of
degenerate models is almost negligible, thus the best fitting
model (which is plotted in the appendix) corresponds quite
closely to an absolute maximum. On the other hand, for those
spectra with strong emission lines degeneracies are more severe,
to such extent that for NGC1569, NGC3310, NGC3690, NGC4194,
NGC4449, NGC4485, NGC4631, NGC4670, NGC4775, NGC5996, NGC6052 and
NGC6240 other acceptable fits that differ significantly from the
best fit presented in the appendix exist. For example, an
acceptable fit is also one where most of the star formation takes
place in the youngest bins and the metallicity is about solar.
This of course would be cured if we were able to have more
detailed absorption features in the spectrum and especially more
resolution, since this would constrain some of the models that
now appear as plausible. Indeed, a larger wavelength coverage
would also help. In any case, it is worth reminding that this
degeneracy is generic of the present problem we are dealing with,
and that using MOPED only facilitates the search in a large
parameter space for such degeneracies and to understand the
physics behind the problem.

The speed up factors obtained in the present analysis are quite
remarkable. On a 1Ghz Athlon machine it takes much less than a
second for an initial starting guess to find a (possibly local)
maximum of the likelihood. For 400 initial guesses it takes about
20 seconds of CPU time to find the global maximum. If one used
the whole data set the same calculation would take about an hour
and a half.  Extrapolating these numbers one can see that for
analysing a big dataset like the 2dF or SDSS, the time required
using all of the data would be about 170 years, longer if
higher-resolution models are used. This is obviously too long to
be achieved, especially if one takes into account that the
theoretical model may change, necessitating a full repeat of the
analysis. With MOPED, it would take a few months on a single
workstation, or much less by exploiting the trivially parallel
nature of the problem.  Finally, the rapid algorithm here has
allowed us to explore more fully the likelihood surface, and we
find that for galaxies with significant very recent star
formation there is often more than one acceptable parameter fit.
There is clearly a need for theoretical stellar atmospheric models
with spectral resolution better than 1\AA\, if we are to extract
all the rich information that large redshift surveys contain in
their galaxy spectra.

\noindent{\bf Acknowledgments}

We are grateful to Karl Glazebrook, Ofer Lahav and Marc
Kamionkowski for useful discussions.


\section*{Appendix I}

The following 12 figures show the best fitting model (thick solid
line) for each spectrum (thin solid line) in the \scite{k92}
atlas (left panel) and the plot of the star formation history for
the 8 age bins (right panel). The SFH is plotted with horizontal
error bars denoting the width of the bins and vertical error bars
for the marginal uncertainty in the height of the bins. Also
shown for each model, is the value of the metallicity derived for
each bin in solar units (left to right correspond to youngest to
oldest bin) and the overall amount of extinction according to the
\scite{C97} model. For those bins with zero star formation the
metallicity is irrelevant and set to zero (and zero error).  Note that
there are five galaxies with Seyfert nuclei: NGC3227, 3516, 5548,
6764, and 7469.

\clearpage
\begin{figure*}
\centerline{\psfig{figure=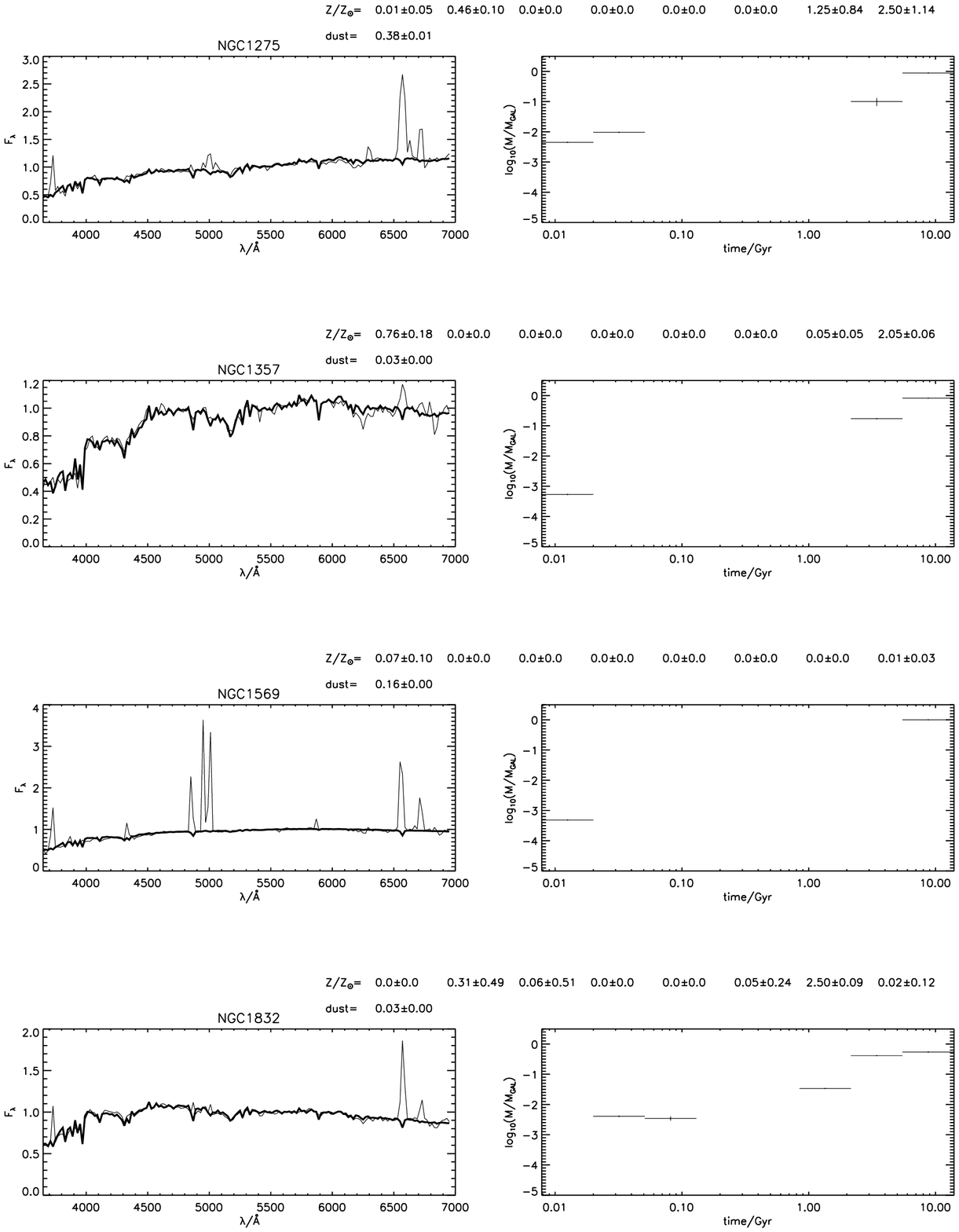,height=22cm,angle=0}}
\caption{}
\end{figure*}
\clearpage

\begin{figure*}
\centerline{ \psfig{figure=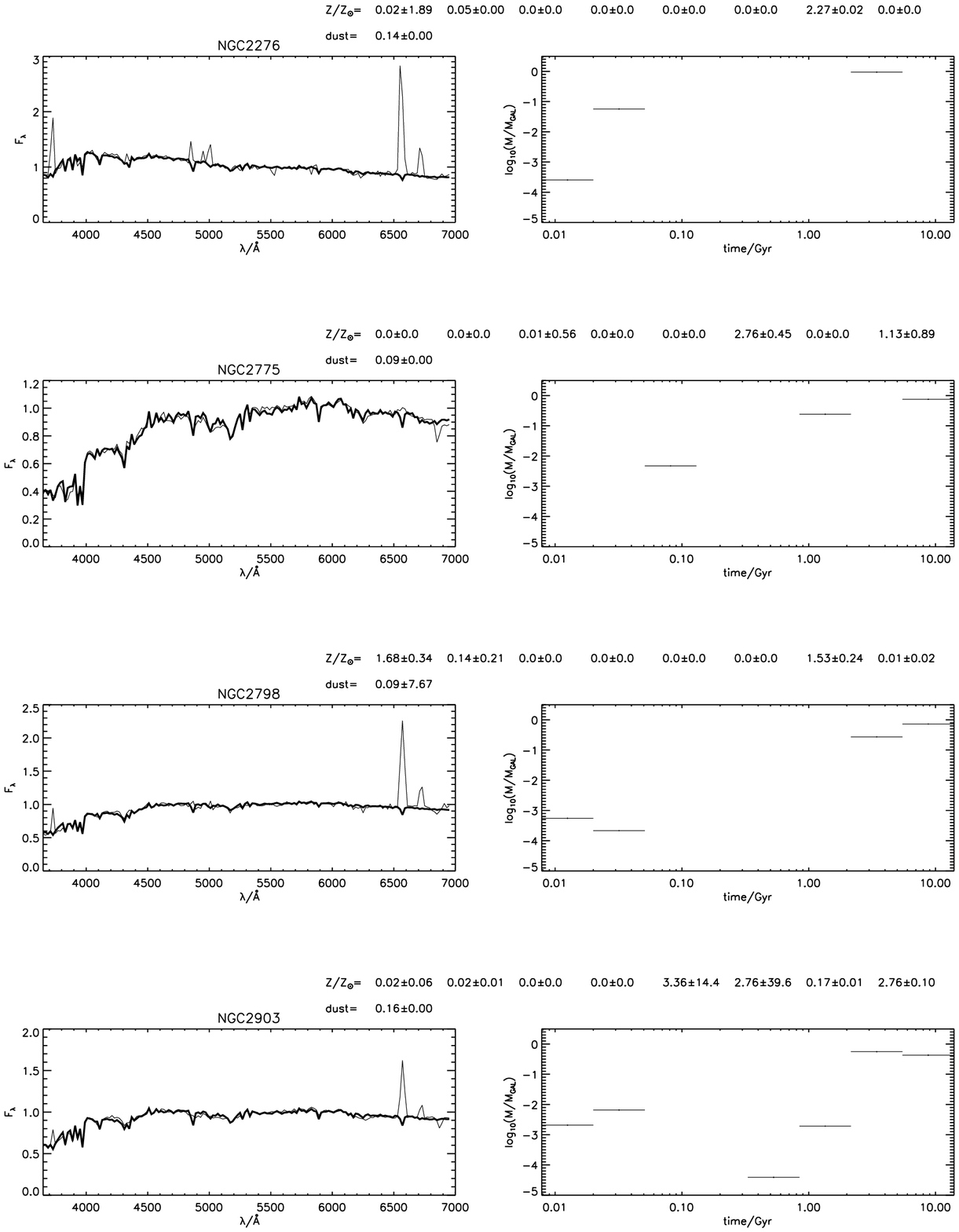,height=22cm,angle=0}}
\caption{}
\end{figure*}
\clearpage

\begin{figure*}
\centerline{ \psfig{figure=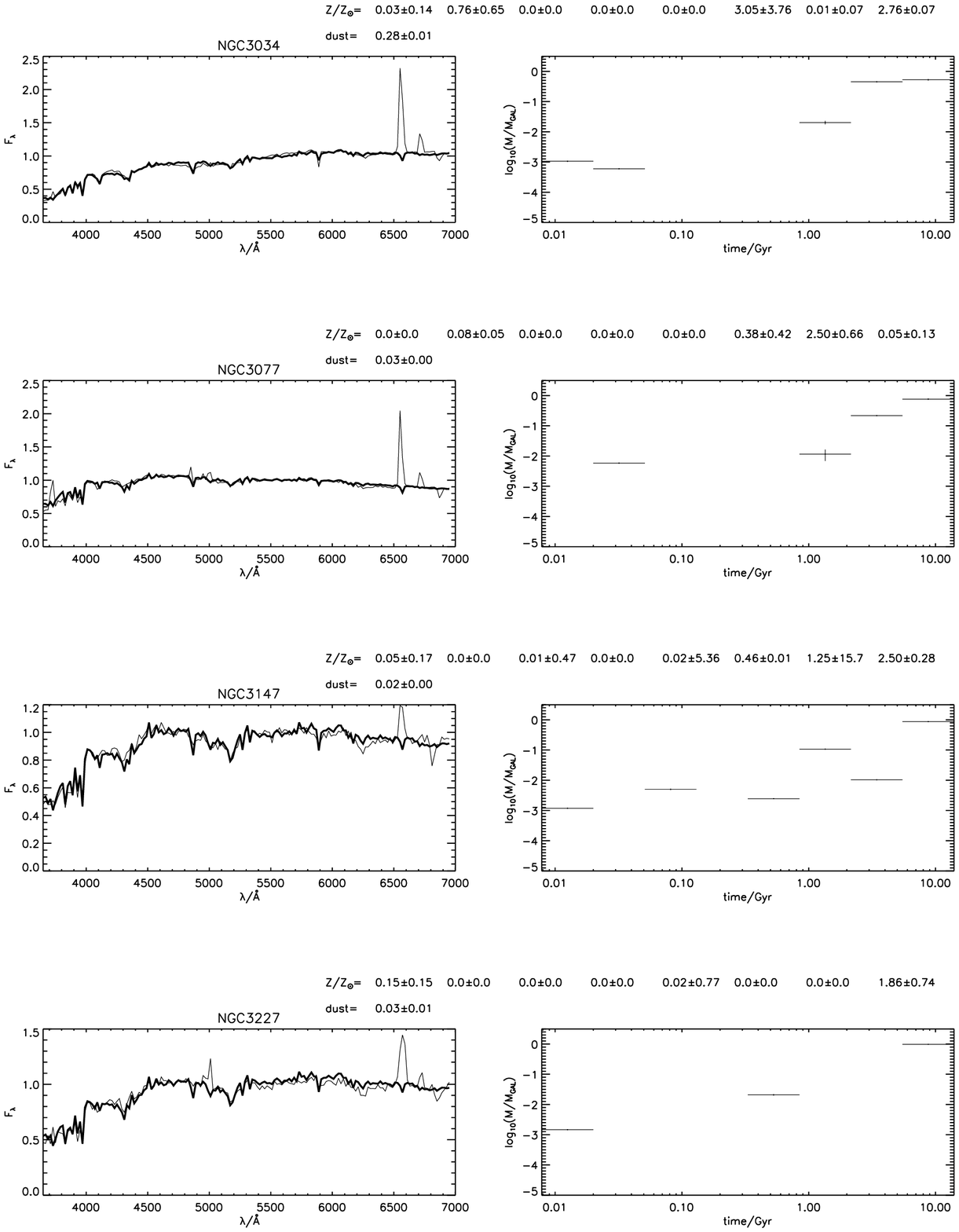,height=22cm,angle=0}}
\caption{}
\end{figure*}
\clearpage

\begin{figure*}
\centerline{ \psfig{figure=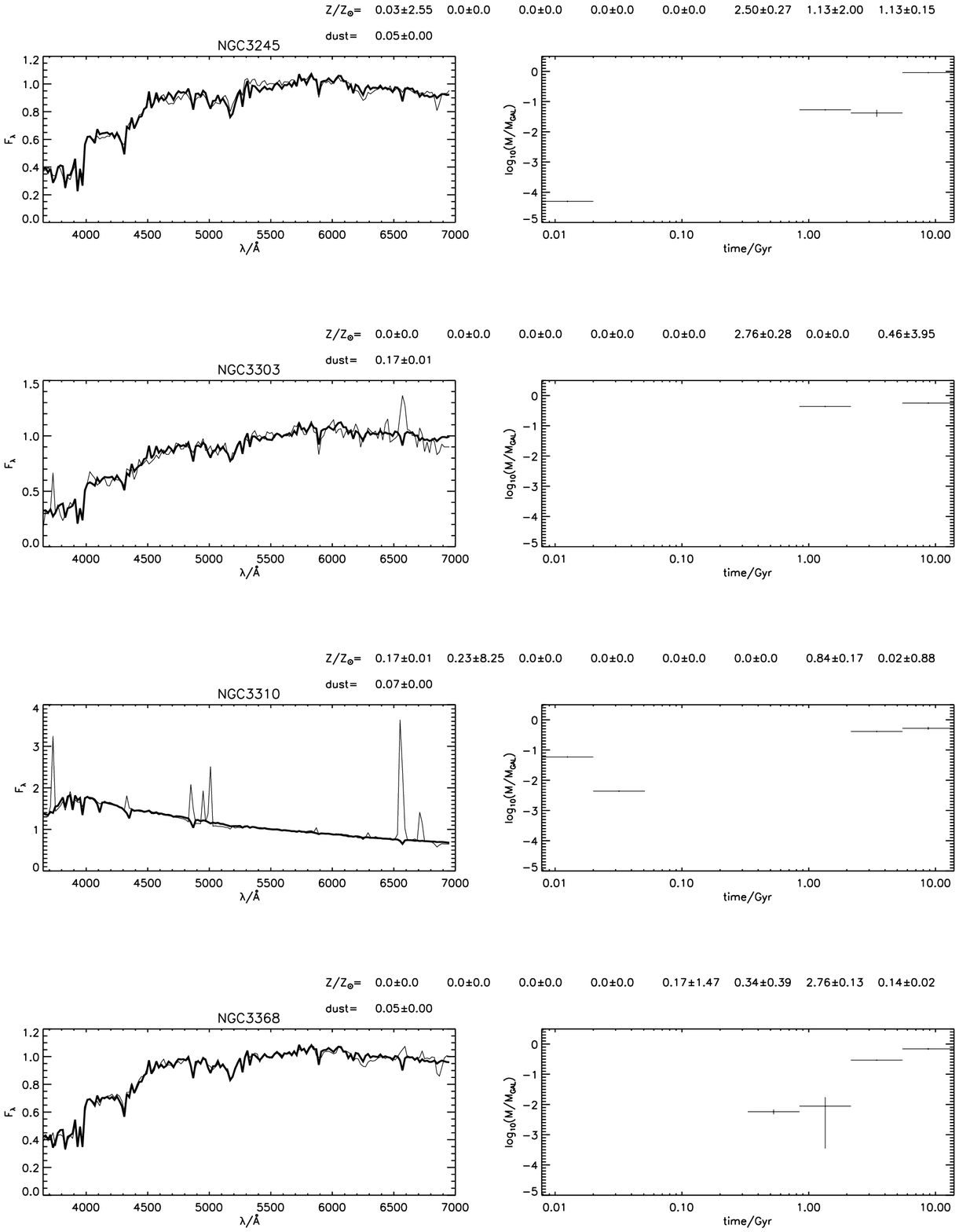,height=22cm,angle=0}}
\caption{}
\end{figure*}
\clearpage

\begin{figure*}
\centerline{ \psfig{figure=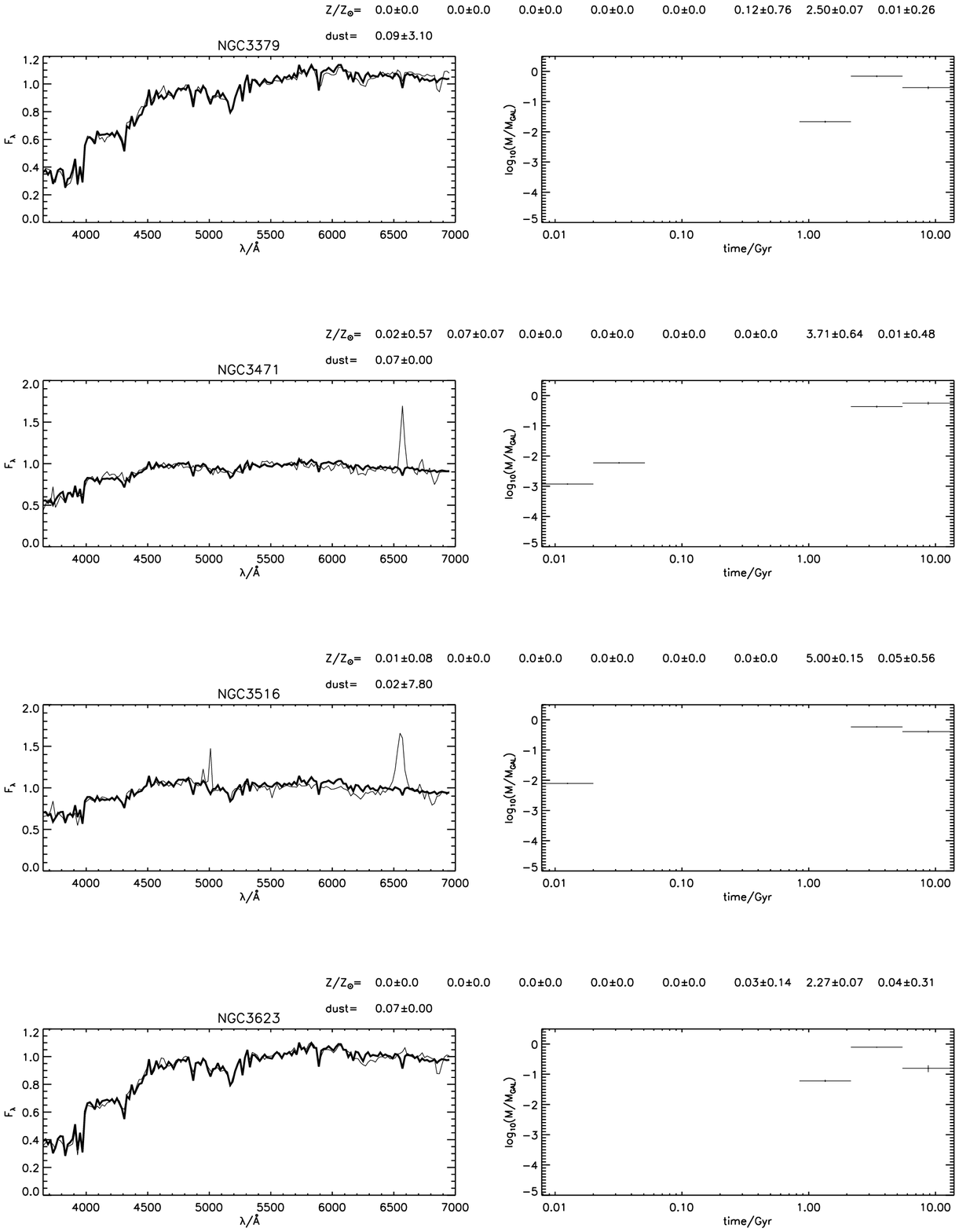,height=22cm,angle=0}}
\caption{}
\end{figure*}
\clearpage

\begin{figure*}
\centerline{ \psfig{figure=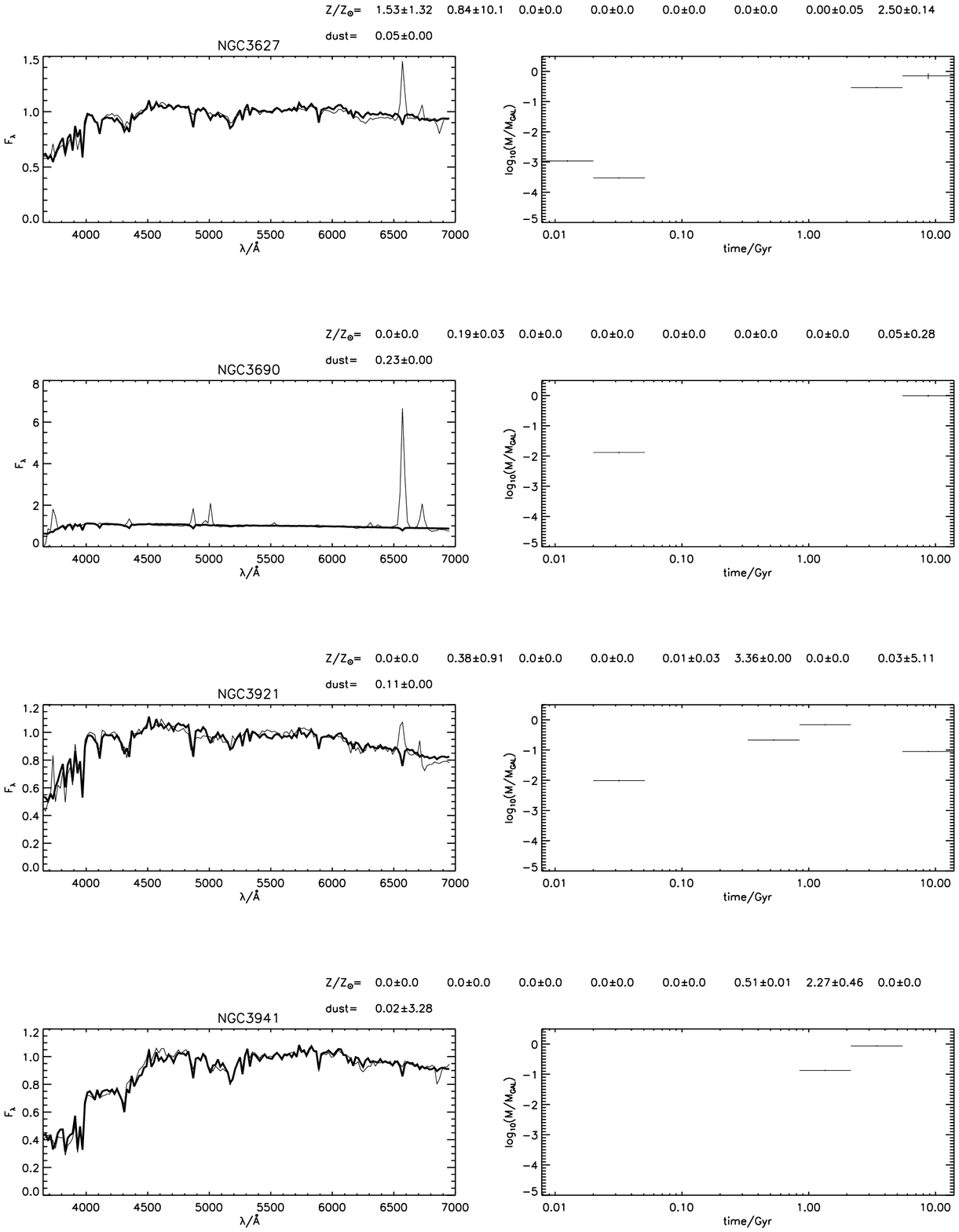,height=22cm,angle=0}}
\caption{}
\end{figure*}
\clearpage

\begin{figure*}
\centerline{ \psfig{figure=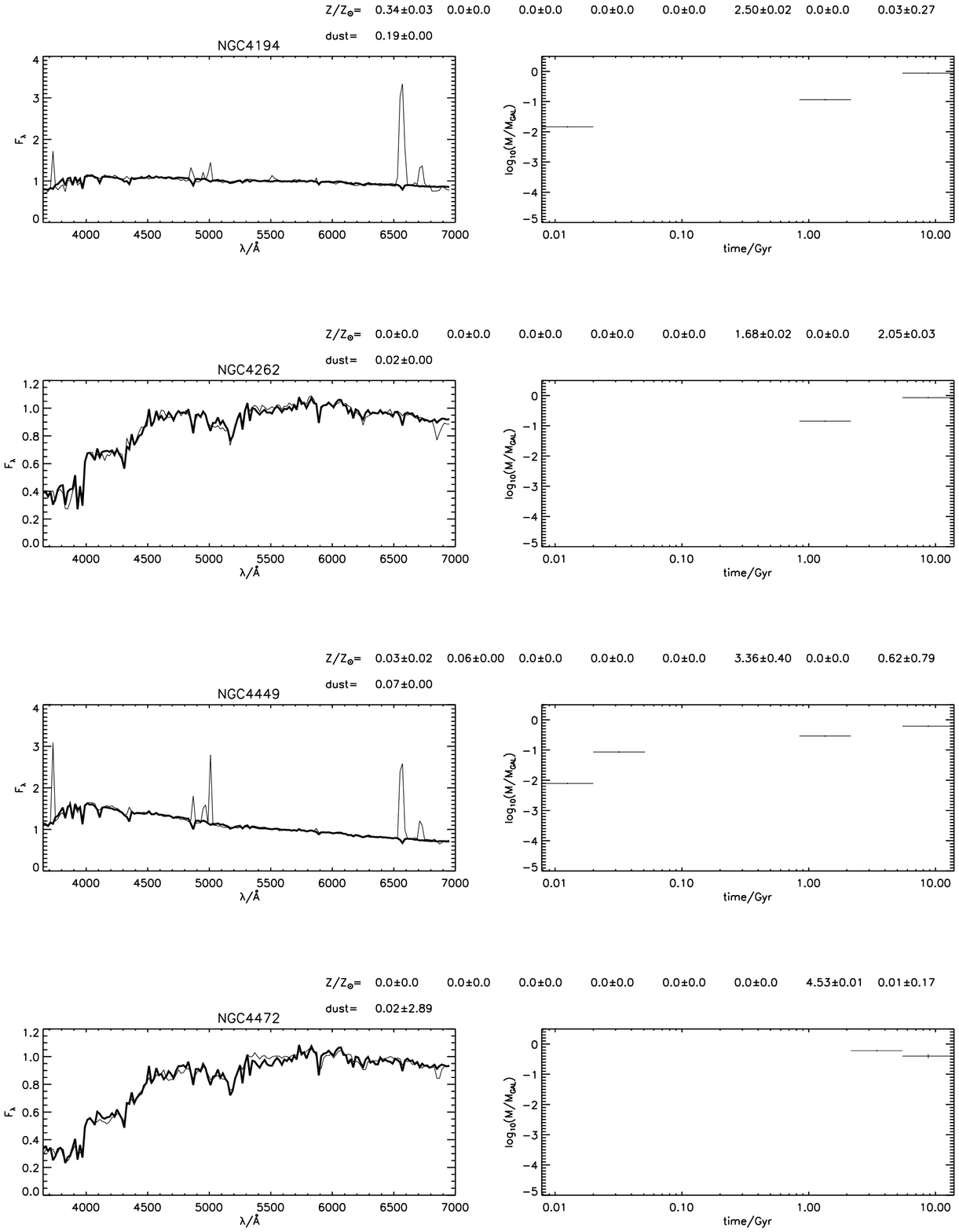,height=22cm,angle=0}}
\caption{}
\end{figure*}
\clearpage

\begin{figure*}
\centerline{ \psfig{figure=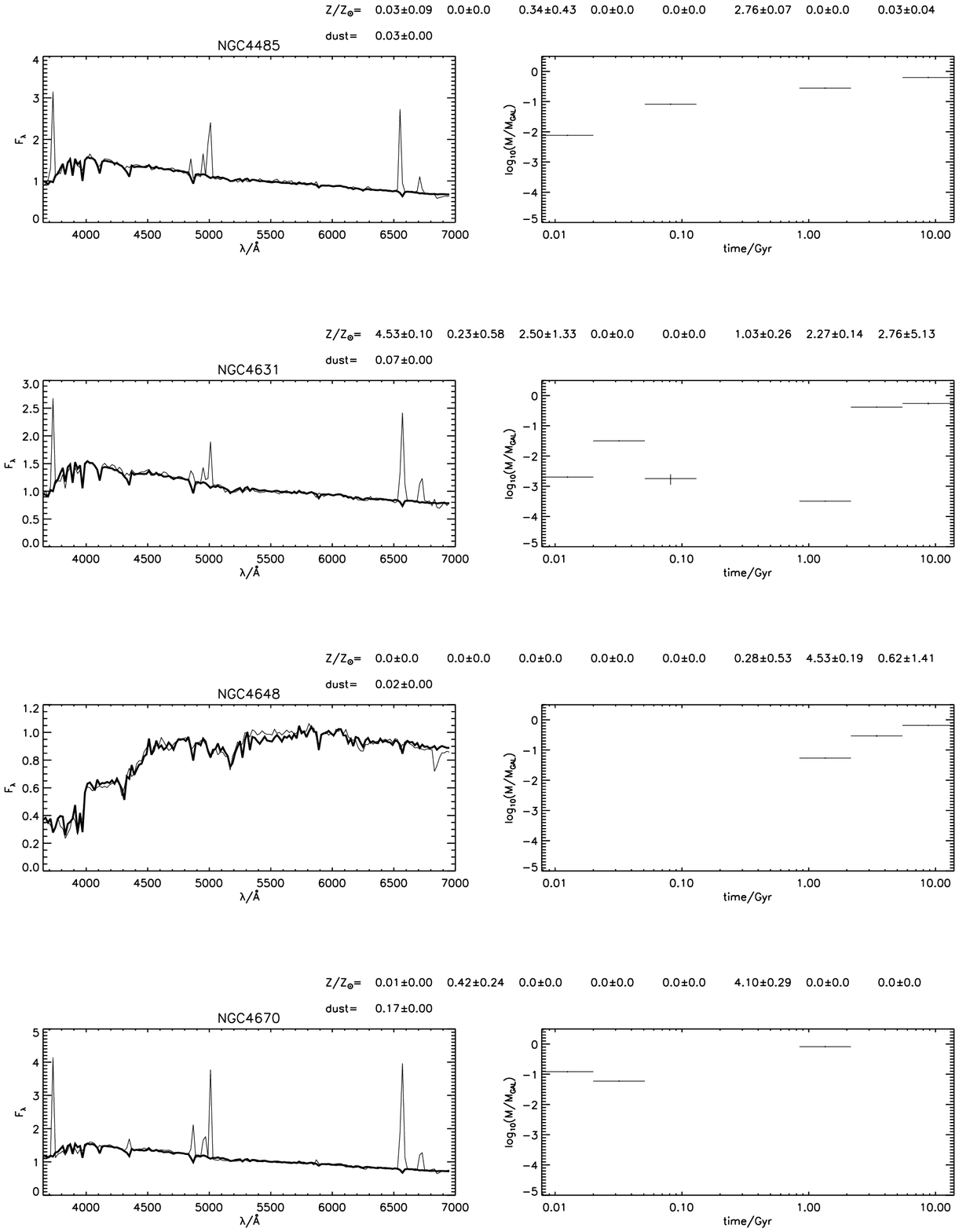,height=22cm,angle=0}}
\caption{}
\end{figure*}
\clearpage

\begin{figure*}
\centerline{ \psfig{figure=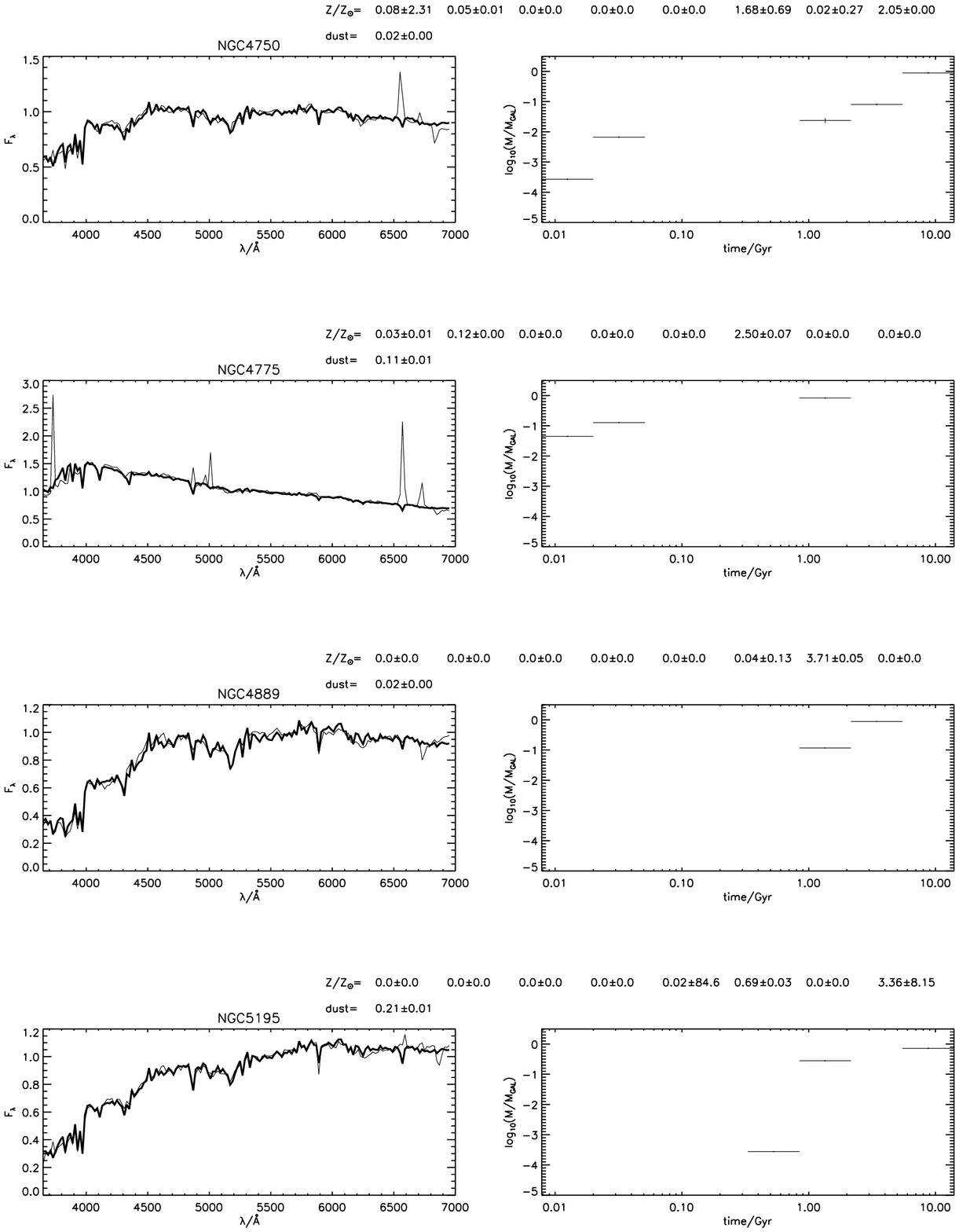,height=22cm,angle=0}}
\caption{}
\end{figure*}
\clearpage

\begin{figure*}
\centerline{ \psfig{figure=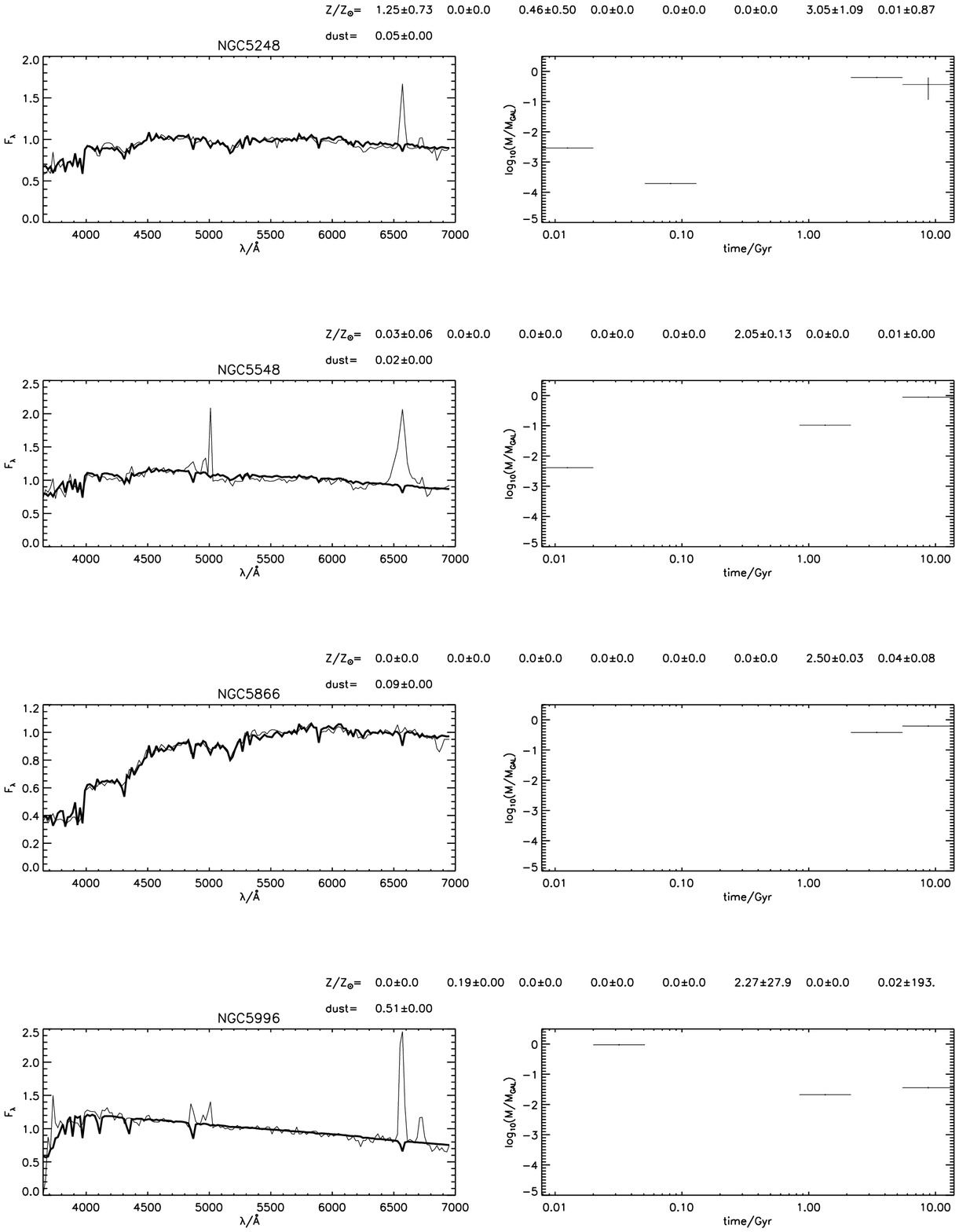,height=22cm,angle=0}}
\caption{}
\end{figure*}
\clearpage

\begin{figure*}
\centerline{ \psfig{figure=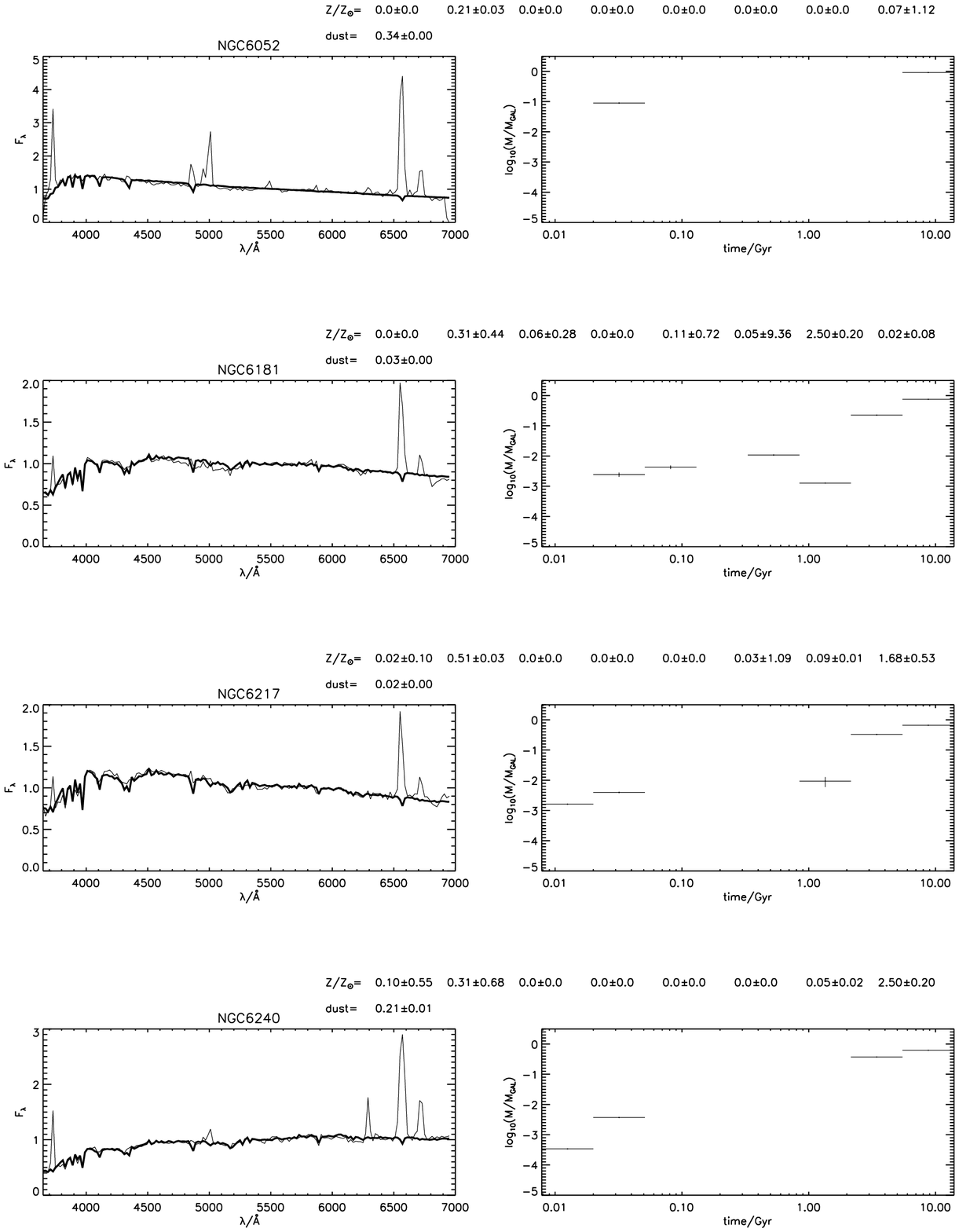,height=22cm,angle=0}}
\caption{}
\end{figure*}
\clearpage

\begin{figure*}
\centerline{ \psfig{figure=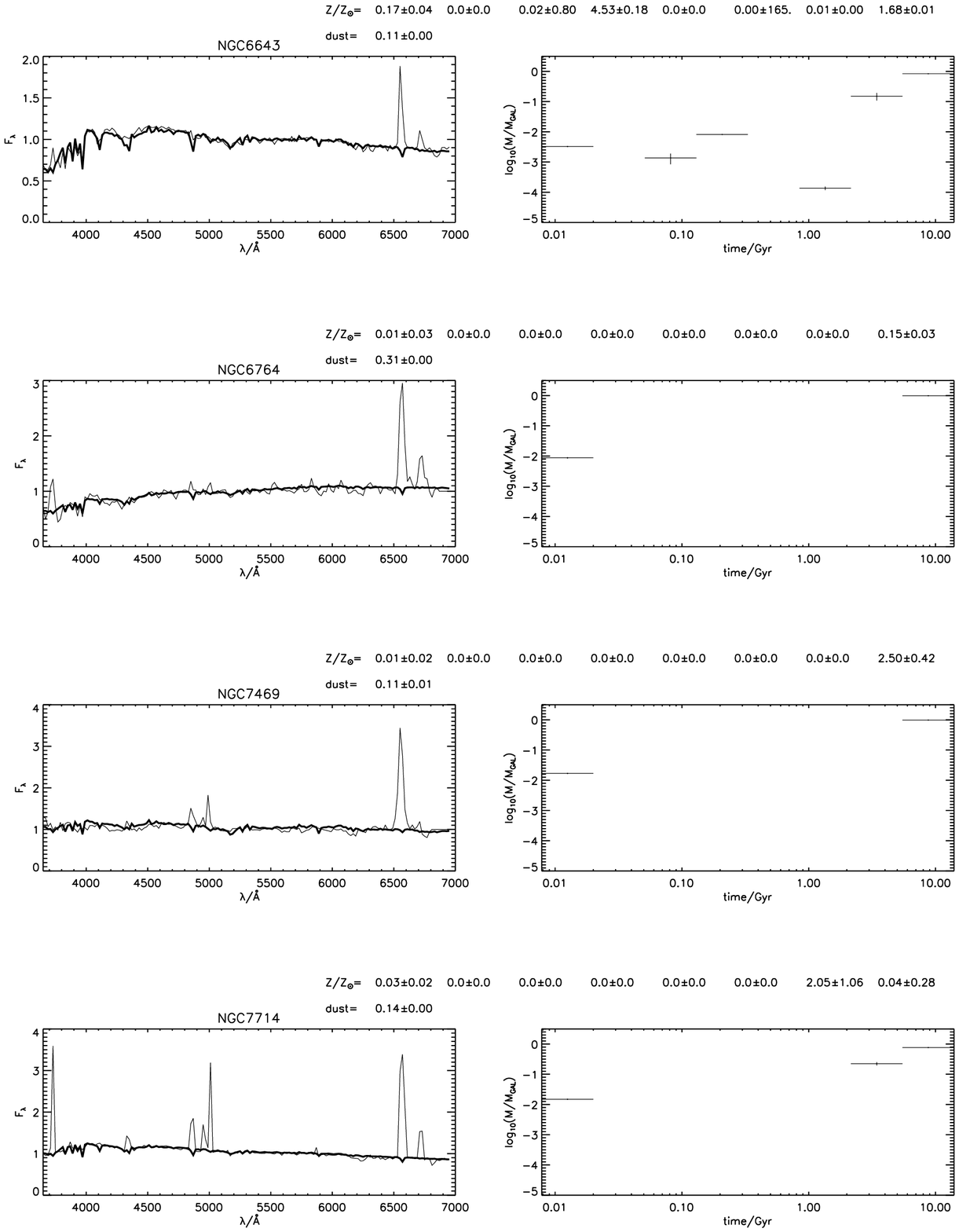,height=22cm,angle=0}}
\caption{}
\end{figure*}
\clearpage

\end{document}